\documentclass[journal=nalefd,manuscript=letter, layout=twocolumn]{achemso}

\usepackage[version=3]{mhchem} 
\usepackage[T1]{fontenc}       
\usepackage{physics}
\usepackage{hyperref}
\usepackage{mathtools, cuted}

\title{Ultra-efficient superconducting Dayem bridge field-effect transistor}
\author{Federico Paolucci}
\affiliation{NEST, Instituto Nanoscienze-CNR and Scuola Normale Superiore, I-56127 Pisa, Italy}
\email{federico.paolucci@nano.cnr.it}
\author {Giorgio De Simoni}
\affiliation{NEST, Instituto Nanoscienze-CNR and Scuola Normale Superiore, I-56127 Pisa, Italy}
\author{Elia Strambini}
\affiliation{NEST, Instituto Nanoscienze-CNR and Scuola Normale Superiore, I-56127 Pisa, Italy}
\author{Paolo Solinas}
\affiliation{SPIN-CNR, Via Dodecaneso 33, I-16146 Genova, Italy}
\author{Francesco Giazotto}
\affiliation{NEST, Instituto Nanoscienze-CNR and Scuola Normale Superiore, I-56127 Pisa, Italy}


\begin{document}
\begin{tocentry}
\center
\includegraphics  [height=3.5cm]{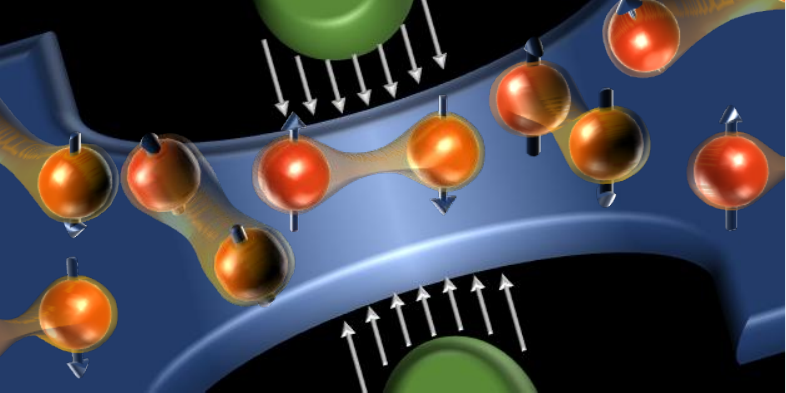}

\end{tocentry}

\begin {strip}
\begin{abstract}

Superconducting field-effect transitor ($SuFET$) and Josephson field-effect transistor ($JoFET$) technologies take advantage of electric field induced control of charge carrier concentration in order to modulate the channel superconducting properties. Despite field-effect is believed to be unaffective for superconducting metals, recent experiments showed electric field dependent modulation of the critical current ($I_C$) in a fully metallic transistor. Yet, the grounding mechanism of this phenomenon is not completely understood. Here, we show the experimental realization of Ti-based Dayem bridge field-effect transistors ($DB-FET$s) able to control $I_C$ of the superconducting channel. Our easy fabrication process $DB-FET$s show symmetric full suppression of $I_C$ for an applied critical gate voltage as low as $V_G^C\simeq \pm8$V at temperatures reaching about the $85\%$ of the record critical temperature $T_C\simeq550$mK for titanium. The gate-independent $T_C$ and normal state resistance ($R_N$) coupled with the increase of resistance in the supercoducting state ($R_S$) for gate voltages close to the critical value ($V_G^C$) suggest the creation of field-effect induced metallic puddles in the superconducting sea. Our devices show extremely high values of transconductance ($|g_m^{MAX}|\simeq15\mu$A/V at $V_G\simeq \pm 6.5$V) and variations of Josephson kinetic inductance ($L_K$) with $V_G$ of two orders of magnitude. Therefore, the $DB-FET$ appears as an ideal candidate for the realization of superconducting electronics, superconducting qubits, tunable interferometers as well as photon detectors.
\end{abstract}
\end{strip}


Conventional computation is hinged on the use of field-effect transistors ($FET$s)\cite{Lilienfeld1926, Nishizawa1982} based on complementary metal-oxide-semiconductor ($CMOS$) technology \cite{Wanlass1967}. Semiconductor-based electronics suffers from dissipation caused by charging and de-charging the gate capacitors, and by the current flowing through the $FET$ channel\cite{Tolpygo2016}. The latter can be eliminated by employing a class of devices where a high critical temperature ($T_C$) superconductor thin film substitutes the semiconducting channel: the superconducting field-effect transistor ($SuFET$) \cite{Nishino1989}. By applying a gate voltage ($V_G$) the charge carrier concentration in the channel can be controlled (because of their low intrinsic carrier density) and, as a consequence, the normal state resistance ($R_N$), the superconducting critical temperature \cite{Fiory1990, Mannhart1993} and the critical current ($I_C$) are modulated \cite{Okamoto1992, Mannhart1993b}. A similar approach consists in employing superconducting source and drain electrodes which induce Cooper pairs flowing in a semiconducting channel through the so-called superconducting proximity effect \cite{Holm1932}. In the resulting device, called Josephson field-effect transistor ($JoFET$)\cite{Clark1979}, $T_C$ and $I_C$ can be controlled via field-effect modulation of the carrier concentration of the semiconductor \cite{Takayanagi1985, Akazaki1996}. Since the first demonstration of Josephson supercurrent in proximized semiconductor nanowires \cite{Doh2005, Xiang2006, Paajaste2015}, 1D channels have been implemented in $JoFET$ technology\cite{Jespersen2009, Abay2014}, too. This gave rise to novel experiments shedding light on long standing physics open problems, such as Majorana bound states \cite{Mourik2012, Das2012}, and developing new superconductor-based qubit technologies \cite{Larsen2015, deLange2015, Casparis2016}. Despite field-effect is believed to be unaffective on conventional superconducting metals, recent experiments showed full suppression of supercurrent in all-metallic transistors based on different Bardeen-Cooper-Schrieffer ($BCS$) wires made of aluminum and titanium \cite{DeSimoni2018}. This discovery could pave the way to the realization of novel all-metallic gate-tunable devices, such as "gatemons" \cite{Larsen2015, deLange2015, Casparis2016}, interferometers\cite{Clarke2004}, Josephson parametric amplifiers\cite{Bergeal2010, Kamal2011} and photon detectors\cite{Goltsman2001}, that could take advantage of a simple single-step fabrication and scalable technology.

Here, we report the realization of the first fully metallic Dayem bridge field-effect transistors ($DB-FET$s). Remarkably, our titanium-based devices show an unprecedented superconducting critical temperature $T_C\sim540$mK for Ti \cite{Peruzzi1999, Tirelli2008, Faivre2008}, and a symmetric complete suppression of the supercurrent for voltages applied to the lateral gate electrodes as large as $\pm 8$V. Furthermore, the $DB-FET$s exhibit values of transconductance reaching $15\mu$A/V at about $\pm 6.5$V and variations of two order of magnitude in the Josephson kinetic inductance ($L_K$) with the applied $V_G$. 

\begin{figure*}[t!]
  \includegraphics [width=8.4cm] {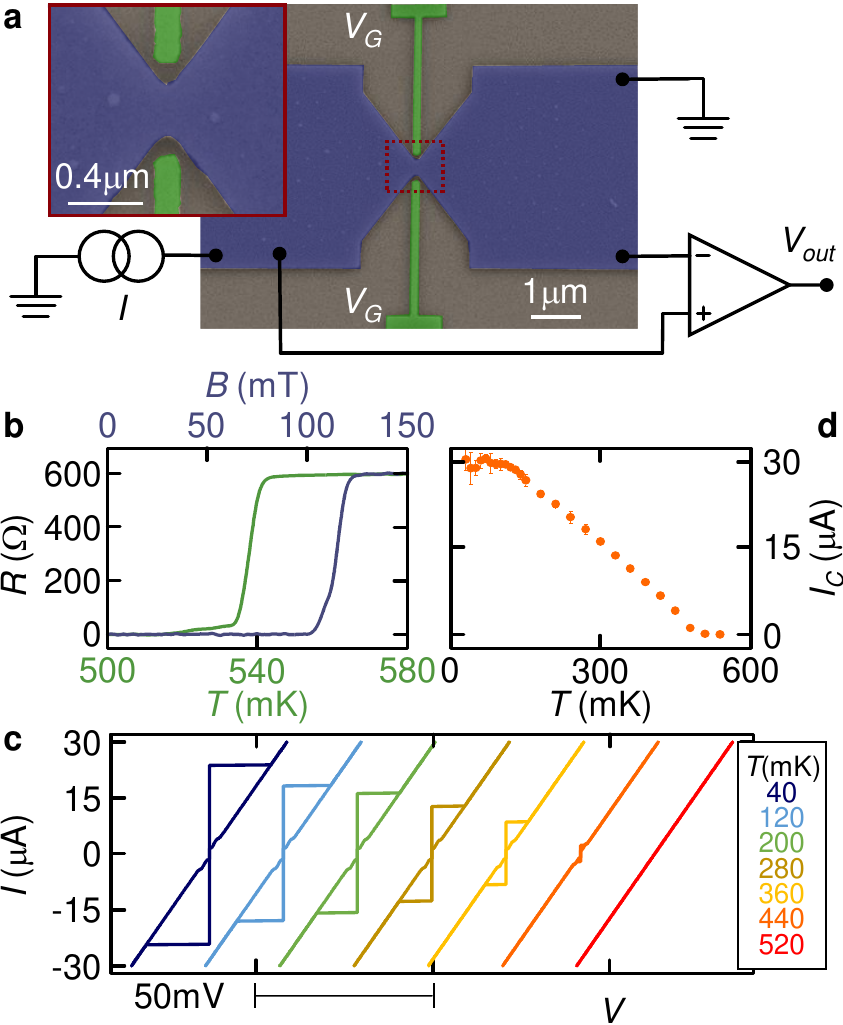}
  \caption{(a) False color electron micrograph of a typical Ti-based Dayem bridge transistor. The Josephson junction (blue) is current biased and the voltage drop is measured with a room temperature voltage amplifer, while the gate voltage is applied to both gate electrodes (green fingers). The inset shows a blow-up of the core of the device. (b) Resistance $R$ as a function of temperature $T$ (green line, bottom horizontal axis), and $R$ vs perpendicular-to-plane magnetic field $B$ at $50$mK (blue line, top horizontal axis) characteristics of the $DB-FET$. The normal-state resistance of the device is about 600$\Omega$. (c) Back and forth current $I$ vs voltage $V$ characteristics of a typical $DB-FET$ measured at different temperatures. The curves are horizontally offset by 20 mV for clarity. (d) Full temperature evolution of the critical current $I_C$. The error bars represent the standard deviation of $I_C$ calculated over 50 repetitions.}
  \label{Figure1}
\end{figure*}

The structure of a typical Dayem bridge field-effect transistor is shown in Figure \ref{Figure1}a. The $DB-FET$ consists of a $4\mu$m wide titanium (Ti) thin film (of thickness $\sim30$nm) interrupted by a $\sim 125$-nm-long and $\sim 300$-nm-wide constriction. In correspondence of the constriction, at a distance of about $120-150$nm, two side electrodes (green stripes in Figure \ref{Figure1}a) allow to apply an electrostatic field on the bridge region. The devices were realized by a single-step electron beam lithography and evaporation of titanium onto a p$^{++}$-doped silicon (Si) commercial wafer covered by $300$nm of silicon dioxide (SiO$_2$). The $30$-nm-thick Ti layers were deposited at room temperature in an ultra-high vacuum electron beam evaporator (base pressure $\sim 10^{-11}$Torr) at a deposition rate ranging from $10$ to $13$\AA/s. The electrical characterization of the $DB-FET$s was performed by standard four wire technique in a filtered He$^3$-He$^4$ dry dilution refrigerator at different bath temperatures (in the range 40mK - 550mK). Both resistance vs temperature and resistance vs magnetic field characteristics were obtained by low frequency lock-in technique, while the current vs voltage behaviors were carried out by applying a low-noise current bias and measuring the voltage drop by a room-temperature differential preamplifier. Finally, the gate voltage was applied by a low-noise source-measurement-unit.

Our Dayem bridges show an unprecedented critical temperature for titanium $T_C\sim540$mK \cite{Peruzzi1999, Tirelli2008, Faivre2008} (see the resistance vs temperature trace depicted by the green line in  Figure \ref{Figure1}b) corresponding to a $BCS$ zero-temperature superconducting energy gap $\Delta_0=1.764k_BT_C\simeq82\mu$eV, where $k_B$ is the Boltzmann constant. On the other hand, our $DB-FET$s exhibit typically a critical perpendicular-to-plane magnetic field of $\sim115$mT, as displayed by the blue line of Figure \ref{Figure1}b. In order to investigate the dissipationless Cooper pairs transport, we measured the $I-V$ characteristics of our devices as a function of bath temperature ($T$). A set of $I-V$ characteristics at selected temperatures is represented in Figure \ref{Figure1}c. The superconducting critical current reaches a value $I_C\simeq30\mu$A almost constant for temperatures lower than $\sim125$mK and monotonically decreases for higher temperature, as shown in Figure \ref{Figure1}d. The transistor $I-V$ characteristics are hysteretic due to heating induced in the sample while switching from the normal to the superconducting state \cite{Courtois2008}. In particular, the switching current from the resistive to the dissipationless state, known as retrapping current, is almost constant ($I_R\simeq1.6\mu$A) in the complete temperature range.

\begin{figure*}[t!]
  \includegraphics [width=8.4cm] {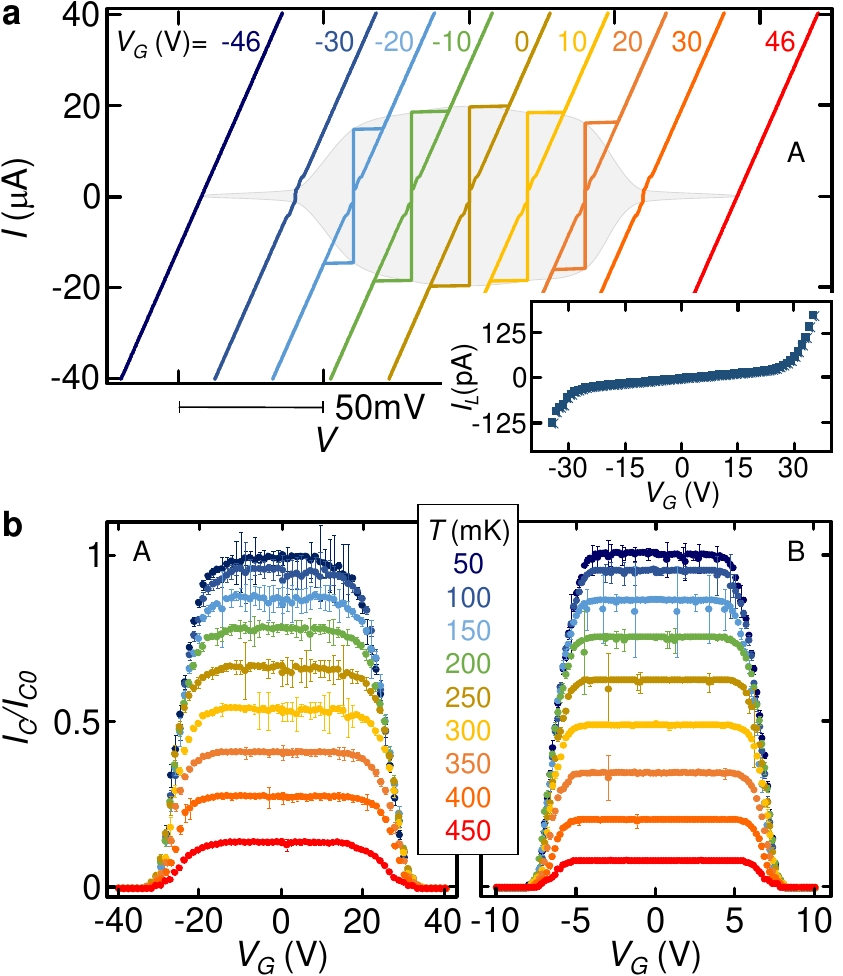}
  \caption{(a) Back and forth current $I$ vs voltage $V$ characteristics of sample $A$ measured at 50mK for several values of gate voltage $V_G$. The curves are horizontally offset proportionally to $V_G$ for clarity. The semi-transparent area depicts the parameters space where superconductivity persists. The inset shows the leackage current measured in our setup $I_L$ as a function of gate voltage $V_G$. (b) Normalized critical current $I_C/I_{C0}$ as a function of gate voltage $V_G$ measured at different bath temperatures $T$ for two $DB-FET$ ($i.e.$, samples $A$ and $B$). The error bars represent the standard deviation of $I_C$ calculated over 50 repetitions.}
  \label{Figure2}
\end{figure*}

To demonstrate the field-effect performances of the $DB-FET$, we carried out $I-V$ measurements for different values of gate voltage. In our experiments we applied the same value of $V_G$ to both gate electrodes to maximize the impact of the electric field on the supercurrent. As shown in  Figure \ref{Figure2}a for sample $A$, the critical current $I_C$ monotonically decreases with the applied gate voltage untill reaching full suppression. On the other hand, the retrapping current ($I_R$) remains constant untill $I_C>I_R$, while by further increasing $V_G$ the switching and retrapping currents assume the same value ($I_R=I_C$). Remarkably, the suppression of $I_C$ is symmetric with the sign of $V_G$ ($bipolar$ field-effect) and the normal state resistance ($R_N$) of the $DB-FET$ is unaffected by the electric field (see the constant slope of the $I-V$ curves in Figure \ref{Figure2}a). This is in stark contrast with $SuFET$s\cite{Nishino1989,Fiory1990, Mannhart1993,Okamoto1992, Mannhart1993b} and $JoFET$s \cite{Takayanagi1985, Akazaki1996, Doh2005, Xiang2006, Paajaste2015}. As a consequence, charge depletion of the Josephson transistor channel can not account for the $I_C$ reduction. 

In order to study the complete beahvior of the $DB-FET$s we measured $I_C$ as a function of the applied gate voltage at different bath temperatures ranging from $50$mK to $450$mK. By normalizing the critical current with respect to its value in the absence of any applied gate bias
 [$I_C(V_G)/I_C(0)$], it is possible to directly compare the behavior of different devices, as shown in  Figure \ref{Figure2}b, even if their intrisic critical current is different [$I_C(T=50\text{mK})\simeq 28\mu$A for sample $A$ and $I_C(T=50\text{mK})\simeq 24\mu$A for sample $B$]. The critical current of the two devices shows a qualitatively similar behavior both with gate voltage and temperature. In particular, at $T=50$mK the critical current remains constant by increasing $V_G$, then it starts to decrease untill its full suppression ($I_C=0$) by further rising the gate voltage. For higher values of the temperature the maximum value of critical current lowers, and, on the one hand, the plateau of constant $I_C$ widens in $V_G$ while, on the other hand, the foot of the transition ($i.e.$, the first value of $V_G$ giving $I_C=0$) remains constant at temperatures up to $\sim85\%T_C$ (see Figure \ref{Figure2}b). This behavior resembles the recent results obtained on gated $BCS$ Ti and Al wires and thin films \cite{DeSimoni2018}. From a quantitative point of view, the main difference between sample $A$ and $B$ resides in the gate voltage operation ranges (see Figure \ref{Figure2}b). Sample $A$ exhibits complete suppression of the critical current at a critical voltage$V_G^C\simeq\pm32$V, whereas sample $B$ shows an extraordinary low critical voltage of about $\pm8$V. The latter impressive value of $V_G^C$ is comparable to the gate voltages employed in $CMOS$ technology ($V_{dd}=5$V), and it can be attribuited to the lower distance between the constriction (active element) and the gate electrodes in sample $B$.

\begin{figure*}[t!]
  \includegraphics [width=8.4cm] {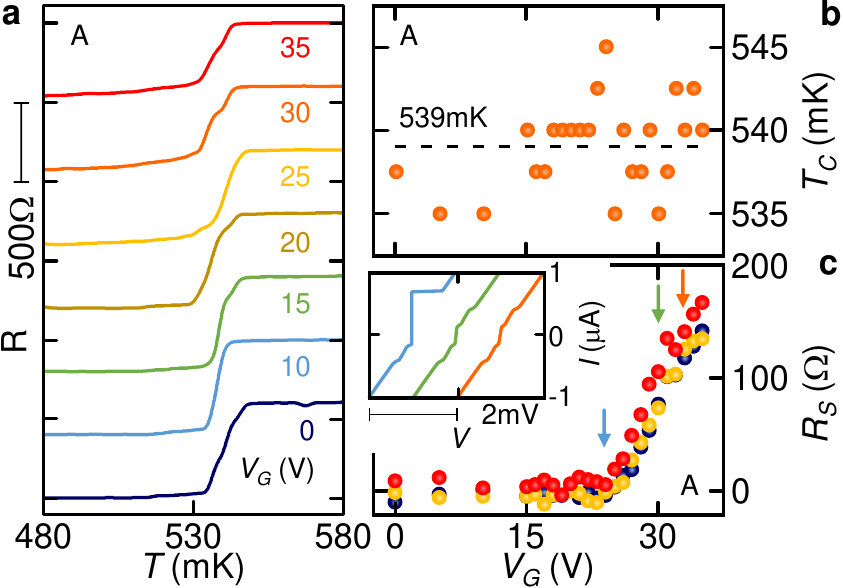}
  \caption{(a) Resistance $R$ as a function of temperature $T$ measured on sample $A$ for different values of gate voltage $V_G$. The curves are vertically offset for clarity. (b) Extracted values of critical temperature $T_C$ as a function of gate voltage $V_G$ for sample $A$. The avarage critical temperature $T_C=539$mK is shown. (c) Zero-bias resistance in the superconducting state $R_S$ as a function of $V_G$ measured on sample $A$ for $T=$ $480$mK (blue), $500$mK (yellow) and $520$mK (red). The inset shows the $I-V$ characteristics at $T_{bath}=50$mK for values of gate voltage indicated by the colored arrows.}
  \label{Figure3}
\end{figure*}

In addition, during all experiments we have carefully monitored the leakage current $I_L$ of the entire measurement circuit while appling the gate voltage $V_G$ (see the inset of Figure \ref{Figure2}a). The leakage current is always of the order of tens of pA ($I_L\sim10^{-6}I_C(V_G=0)$). In particular, for $V_G=25$V the critical current suppression is $\sim13\mu$A while $I_L\simeq20$pA (with a resulting gate resistance $R_G\simeq1.25$T$\Omega$). In order to exclude any quasiparticle overheating due to direct injection of a portion of $I_L$ as source of the supercurrent suppression we recorded the temperature dependence of the resistance ($R$) of our $DB-FET$s as a function of the applied gate voltage (see Figure \ref{Figure3}a). Any heat injection able to suppress the critical current would, at the same time, result in a measurable reduction of the critical temperature \cite{Morpurgo1998}. In our Dayem bridge transistors, $T_C$ remains constant (within the experimental error due to temperature stabilization) over the complete range of applied gate voltages, as shown for sample $A$ in Figure \ref{Figure3}b. For instance, by applying a gate voltage $V_G=25$V a reduction of $80\%$ in $I_C$ is measured while the critical temperature is completely unaffected. Analogously, the normal state resistance is not influenced by the gate voltage and shows an electric field independent value $R_N\simeq600\Omega$ comparable with the values extracted from the $I-V$ characteristics.

A carefull analysis of the $R$ vs $T$ data highlights that the resistance in the superconducting state ($R_S$) near full suppression of the supercurrent strongly depends on $V_G$ , as shown in Figure \ref{Figure3}c. Specifically, a resistive component in the $I-V$ characteristics around zero bias
 appears and grows by rising the gate voltage. This phenomenology seems compatible with the creation of an inhomogeneous mixed state composed of normal metal and superconducting puddles. In this view, the superconducting areas maintain the same critical temperature of pristine Ti, while the dissipative parts interrup the supercurrent flow providing the measured resistive component. The latter can be also noticed by zooming the $I-V$ curves obtained for values of gate voltage approaching $V_G^C$: a finite slope builds up ($i.e.$, a finite conductance, see the inset of Figure \ref{Figure3}c) in correspondence of the dissipative component revealed by the $R$ vs $T$ experiments.

\begin{figure*}[t!]
  \includegraphics [width=8.4cm] {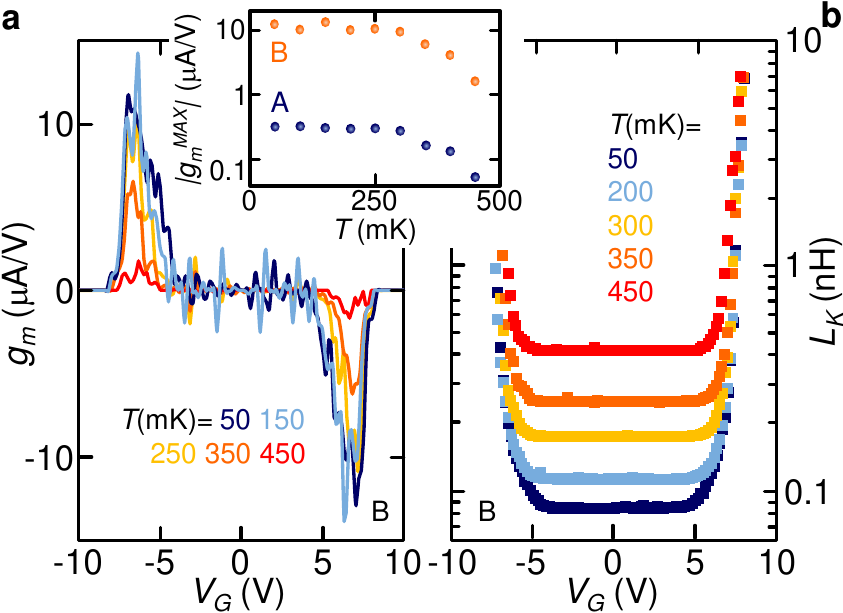}
  \caption{(a) Transconductance $g_m$ as a function of gate voltage $V_G$ for different values of temperature $T$ measured on sample B. The inset shows the absolute value of the maximum of transcoductance $|g_m^{MAX}|$ vs temperature $T$ for devices $A$ and $B$. (b) Kinetic inductance $L_K$ as a function of gate voltage $V_G$ for different values of temperature $T$ measured on sample $B$.}
  \label{Figure4}
\end{figure*}

The standard figure of merit providing information about the performances of field-effect transistors is the transconductance. In superconductor-based devices, it is defined as $g_m=dI_C/dV_G$, $i.e.$ the variation of the critical current with gate voltage. Figure \ref{Figure4}a shows the behavior of $g_m$ as a function of $V_G$ measured at different temperatures for sample $A$. As expected, the transconductance shows different sign for $V_G<0$ and $V_G>0$ (given the suppression of $I_C$ for both gate voltage polarities) since the numerical derivative is performed from negative to positive values of $V_G$. The strong temperature dependence of $g_m$ is highlighted by plotting the absolute value of its maximum $|g_m^{MAX}|$ vs $T$, as depicted in the inset of Figure \ref{Figure4}a for both sample $A$ and $B$. In particular, $g_m^{MAX}$ has an almost constant value ($\sim330$nA/V at $V_G\simeq  \pm 23$V for sample $A$ and $\sim15\mu$A/V at $V_G\simeq \pm 6.5$V for sample $B$) up to $T\simeq T_C /2$, and by further rising the temperature it decreases by about one order of magnitude. The maximum value of the transconductance of our $DB-FET$s is comparable to that obtained on InAs thin films based $JoFET$s ($14\mu$A/V at $V_G=0.7$V \cite{Akazaki1996}) and several orders of magnitude larger than that of semiconductor nanowire-based devices (a few nS  at values of $V_G$ of the order of several tens of volt \cite{Doh2005}). 

In our $DB-FET$s, the gate-dependent suppression of the critical current yields to an increase of the Josephson kinetic inductance, defined as $L_K=\hbar/(2eI_C)$ \cite{Giazotto2008} where $\hbar$ is the reduced Planck constant and $e$ is the electron charge, as shown in Figure \ref{Figure4}b for different temperatures. In particular, the maximum value of $L_K$ in our transistors is $sim8$nH, while its variation ranges from $\sim10^{-13}$H to $\sim10^{-8}$H. Therefore, our $DB-FET$s seem suitable candidates for the realization of fully metallic superconducting qubits\cite{Larsen2015, deLange2015, Casparis2016} and Josephson parametric amplifiers\cite{Bergeal2010, Kamal2011}.

In conclusion, we have realized the first fully metallic Dayem bridge Josephson field-effect transistors allowing to control the critical current by applying a gate voltage. The simple fabrication single-step  process  elects the $DB-FET$ as ideal candidate for a number of technological applications. Notably, our Ti thin films showed a record critical temperature $T_C\sim540$mK for this superconducting metal. On the one hand, the $DB-FET$s allow a bipolar fine control of the critical current untill its full suppression for values of gate voltage of $\sim \pm 8$V. On the other hand, differently from conventional $JoFET$s, the critical temperature and the normal state resistance of the transistors are not affected by an electric field. The constant $T_C$ and the change of $R_S$ with $V_G$ suggest the creation of electric field induced inhomogeneous state in the superconductor consisting of metallic puddles. Furthermore, the $DB-FET$s exhibit very high values of transconductance ($|g_m^{MAX}|\simeq15\mu$A/V at $V_G\simeq \pm 6.5$V) and variations of Josephson kinetic inductance of two orders of magnitude. Therefore, the metallic Dayem bridge Josephson field-effect transistors are excellent candidates for the realization of "gatemons" \cite{Larsen2015, deLange2015, Casparis2016}, parametric amplifiers\cite{Bergeal2010, Kamal2011}, tunable interferometers\cite{Clarke2004} and photon detectors\cite{Goltsman2001}.

\section{Author Contributions}
F.P. and G.D.S. fabricated the samples and performed the experiments. F.P. and G.D.S. analysed the data with inputs from E.S., P.S. and F.G. F.G. conceived the the experiment. F.P. wrote the manuscript with inputs from all the authors. All the authors discussed the results and their implications equally.

\begin{acknowledgement}
The authors acknowledge the European Research Council under the European Unions Seventh Framework Programme (FP7/2007-2013)/ERC Grant No. 615187 - COMANCHE, the European Union (FP7/2007-2013)/REA Grant No. 630925 - COHEAT and the MIUR under the FIRB2013 Project Coca (Grant No. RBFR1379UX) for partial financial support. The work of G.D.S. and F.P. was funded by Tuscany Region under the FARFAS 2014 project SCIADRO. The work of E.S. is funded by a Marie Curie Individual Fellowship (MSCA-IFEF-ST No. 660532-SuperMag).
\end{acknowledgement}


\begin{thebibliography}{99}

\bibitem{Lilienfeld1926}
Lilienfeld, J.~E. Method and apparatus for controlling electric currents. U.S. Patent 1,745,175A, Oct. 8, 1926.

\bibitem{Nishizawa1982}
Nishizawa, J-I. Junction Filed-Effect Devices. In \textit{Semiconductor Devices for Power Conditioning}; Sittig, R., Roggwiller, P., Eds.; Earlier Brown Boveri Symposia; Springer, Boston, MA, 1982;  pp 241-272.

\bibitem{Wanlass1967}
Wanlass, F. Low stad-by power complementary field effect circuitry. U.S. Patent 3,356,858, Dec. 5, 1967.

\bibitem{Tolpygo2016}
Tolpygo, S.~K.
\textit{Low Temp. Phys.} {\bf{2016}}, 42,  361.

\bibitem{Nishino1989}
Nishino, T.; Hatano, M.; Hasegawa, H.; Murai, F.; Kure, T.; Hiraiwa, A.; Yagi, K.; Kawabe, U.
\textit{IEEE Trans. Electron Devices} {\bf{1989}}, 10,  61.

\bibitem{Fiory1990}
Fiory, A.~T.; Herbard, A.~F.; Eick, R.~H.; Mankiewich, P.~M.; Howard, R.~E.; O'Malley, M.~L.
\textit{Phys. Rev. Lett.} {\bf{1990}}, 65,  3441.

\bibitem{Mannhart1993}
Mannhart, J.; Str\"obel, J.; Bednorz, J.~G.; Gerber, Ch.
\textit{Appl. Phys. Lett} {\bf{1993}}, 62,  630.

\bibitem{Okamoto1992}
Okamoto, M.
\textit{IEEE Trans. Electron Devices} {\bf{1992}}, 39,  1661.

\bibitem{Mannhart1993b}
Mannhart, J.; Bednorz, J.~G.; M\"uller, K.~A.; Schlom, D.~G.; Str\"obel, J.
\textit{J. Alloys Compd.} {\bf{1993}}, 195,  519.

\bibitem{Holm1932}
Holm, R; Meissner, W. 
\textit{Z. Phys.} {\bf{1932}}, 74,  715.

\bibitem{Clark1979}
Clark, T.~D.; Prance, R.~J.; Grassie, A.~D.~C. 
\textit{J. Appl. Phys.} {\bf{1980}}, 51,  2739.

\bibitem{Takayanagi1985}
Takayanagi, H.; Kawakami, T.
\textit{Phys. Rev. Lett.} {\bf{1985}}, 54,  2449.

\bibitem{Akazaki1996}
Akazaki, T.; Takayanagi, H.; Nitta, J.; Enoki, T.
\textit{Appl. Phys. Lett.} {\bf{1996}}, 68, 418.

\bibitem{Doh2005}
Doh, Y-J.; van Dam, J.~A.; Roest, A.~L.; Bakkers, E.~P.~A.~M.; Kouwenhoven, L.~P.; De Franceschi, S.
\textit{Science} {\bf{2005}}, 309, 272.

\bibitem{Xiang2006}
Xiang, J.; Vidan, A.; Tinkham, M.; Westervelt, R.~M.; Lieber, C.~M.
\textit{Nat. Nanotechnol.} {\bf{2006}}, 1, 208.

\bibitem{Paajaste2015}
Paajaste, J.; Amado, M.; Roddaro, S.; Bergeret, F.~S.; Ercolani, D.; Sorba, L.; Giazotto, F.
\textit{Nano Lett.} {\bf{2015}}, 15, 1803.

\bibitem{Jespersen2009}
Jespersen, T.~S.; Polianski, M.~L.; S\o rensen, C.~B.; Flensberg, K.; Nyg\aa rd, J.
\textit{New J. Phys.} {\bf{2009}}, 11, 113025.

\bibitem{Abay2014}
Abay, S.; Persson, D.; Nilsson, H.; Wu, F.; Xu, H.~Q.; Fogelstr\"om, M.; Shumeiko, V.; Delsing, P.
\textit{Phys. Rev. B} {\bf{2014}}, 89, 214508.

\bibitem{Mourik2012}
Mourik, V.; Zuo, K.; Frolov, S.~M.; Plissard, S.~R.; Bakkers, E.~P.~A.~M.; Kouwenhoven, L.~P.
\textit{Science} {\bf{2012}}, 336, 1003.

\bibitem{Das2012}
Das, A.; Ronen, Y.; Most, Y.; Oreg, Y.; Heiblum, M.; Shtrikman, H.
\textit{Nat. Phys.} {\bf{2012}}, 8, 887.

\bibitem{Larsen2015}
Larsen, T.~W.; Petersson, K.~D.; Kuemmeth, F.; Jespersen, T.~S.; Krogstrup, P.;  Nyg\aa rd, J.; Marcus, C.~M.
\textit{Phys. Rev. Lett.} {\bf{2015}}, 115,  127001.

\bibitem{deLange2015}
de Lange, G.; van Heck, B.; Bruno, A.; van Woerkom, D.~J.; Geresdi, A.; Plissard, S.~R.; Bakkers, E.~P.~A.~M.; Akhmerov, A.~R.; DiCarlo, L.
\textit{Phys. Rev. Lett.} {\bf{2015}}, 115,  127002.

\bibitem{Casparis2016}
Casparis, L.; Larsen, T.~W.; Olsen, M.~S.;  Kuemmeth, F.; Krogstrup, P.;  Nyg\aa rd, J.; Petersson, K.~D.; Marcus, C.~M.
\textit{Phys. Rev. Lett.} {\bf{2016}}, 116,  150505.

\bibitem{DeSimoni2018}
De Simoni, G.; Paolucci, F.; Solinas, P.; Strambini, E.; Giazotto, F.
\textit{ arXiv:1710.02400} {\bf{2018}}.

\bibitem{Clarke2004}
Clarke, J.; Braginski, J. \textit{The SQUID Handbook} VCH: New York, 2004.

\bibitem{Bergeal2010}
Bergeal, N.; Schackert, F.; Metcalfe, M.; Vijay, R.; Manucharyan, V.~E.; Frunzio, L.; Prober, D.~E.; Schoelkopf, R.~J.; Girvin, S.~M.; Devoret, M.~H.
\textit{Nature} {\bf{2010}},465, 64.

\bibitem{Kamal2011}
Kamal, A.; Clarke, J.; Devoret, M.~H.
\textit{Nat. Phys.} {\bf{2011}},78, 311.

\bibitem{Goltsman2001}
Gol'tsman, G.~N.; Okunev, O.; Chulkova, G.; Semenov, A.; Smirnov, K.; Voronov, B.; Dzardanov, A.
\textit{Appl. Phys. Lett.} {\bf{2001}}, 79, 705.

\bibitem{Peruzzi1999}
Peruzzi, A.; Gottardi, E.; Peroni, I.; Ponti, G.; Ventura, G.
\textit{Nucl. Phys. B} {\bf{1999}}, 78, 576.

\bibitem{Tirelli2008}
Tirelli, S.; Savin, A.~M.; Pascual Garcia, C.; Pekola, J.~P.; Beltram, F.; Giazotto, F.
\textit{Phys. Rev. Lett.} {\bf{2008}}, 101, 077004.

\bibitem{Faivre2008}
Faivre, T.; Golubev, D.; Pekola, J.~P.
\textit{J. Appl. Phys.} {\bf{2008}}, 116, 094302.

\bibitem{Courtois2008}
Courtois, H.; Meschke, M.; Peltonen, J.~T.; Pekola, J.~P.
\textit{Phys. Rev. Lett.} {\bf{2008}}, 101, 067002.

\bibitem{Morpurgo1998}
Morpurgo, A.; Klapwijk, T.~M.; van Wees, B.~J.
\textit{Appl. Phys. Lett.} {\bf{1998}}, 72, 966.

\bibitem{Giazotto2008}
Giazotto, F.; Heikkil\"a, T.~T.; Pepe, G.~P.; Helist\"o, P.; Luukanen, A.; Pekola, J.~P.
\textit{Appl. Phys. Lett.} {\bf{2008}}, 92, 162507.



\end{thebibliography}
\end{document}